\documentclass[traditabstract]{aa}
\usepackage[latin9]{inputenc}
\usepackage{color}
\usepackage{verbatim}
\usepackage{amstext}
\usepackage{amssymb}
\usepackage{graphicx}
\usepackage[authoryear]{natbib}
\PassOptionsToPackage{normalem}{ulem}
\usepackage{ulem}

\makeatletter

\providecommand{\tabularnewline}{\\}

\usepackage{txfonts}\usepackage{longtable}\usepackage{colortbl}
\makeatother

\begin{document}

\title{On the \ion{H}{i}-Hole and AGB Stellar Population of the Sagittarius
Dwarf Irregular Galaxy }

\subtitle{HST Proper-Motion Decontamination%
\thanks{Based on HST programs GO-9820 and GO-10472%
}}

\author{Y. Momany\inst{1,2}\and M. Clemens \inst{2} \and L.R. Bedin\inst{2}
\and M. Gullieuszik\inst{2} \and E.V. Held\inst{2} \and I. Saviane\inst{1}
\and S. Zaggia\inst{2} \and L. Monaco\inst{1} \and M. Montalto\inst{3}
\and R. M. Rich\inst{4} \and L. Rizzi\inst{5} }

\institute{European Southern Observatory, Alonso de Cordova 3107, Santiago,
Chile. \email{ymomany@eso.org} \and INAF, Osservatorio Astronomico
di Padova, Vicolo dell'Osservatorio 5, I-35122 Padova, Italy \and
Centro de Astrofísica, Universidade do Porto, Rua das Estrelas, 4150-762,
Porto, Portugal \and Division of Astronomy, University of California,
8979 Math Sciences, Los Angeles, CA 90095-1562, USA \and W. M. Keck
Observatory, 65-1120 Mamalahoa Highway, Waimea, HI 96743, USA }

\date{Received April 24, 2014; accepted ..., 2014}

\abstract{Using two  HST/ACS data-sets  that are separated  by $\sim2$
  years has  allowed us to  derive the relative proper-motion  for the
  Sagittarius dwarf irregular (SagDIG) and reduce the heavy foreground
  Galactic  contamination.  The  proper-motion  decontaminated  SagDIG
  catalog provides a much clearer view of the young red-supergiant and
  intermediate-age  asymptotic  giant  branch populations.  Previously
  identified carbon and oxygen-rich star samples, based on narrow-band
  filter  photometry, were complemented  with membership  criteria. We
  report the identification of  $3$ Milky Way carbon-rich dwarf stars,
  probably  belonging to  the  thin  disk, and  pointing  to the  high
  incidence of  this class at  low Galactic latitudes. A  sub-group of
  $4$ oxygen-rich  candidate stars depicts  a faint, red  extension of
  the well-defined  SagDIG carbon-rich  sequence. The origin  of these
  oxygen-rich   candidate  stars   remains  unclear,   reflecting  the
  uncertainty in the ratio of carbon/oxygen rich stars.
 Lastly, SagDIG  is a  gas-rich  galaxy characterized  by a  {\em
    single} large  cavity in the gas disk  (\ion{H}{i}-hole), which is
  offset by $\sim360$ pc from  the optical centre of the galaxy.  
  We  nonetheless  investigate  the  stellar  feedback  hypothesis  by
  comparing the  proper-motion cleaned stellar  populations within the
  \ion{H}{i}-hole  with  appropriately  selected  comparison  regions,
  having higher \ion{H}{i} densities external to the hole.
The comparison shows no significant differences.
In particular,  the centre of the \ion{H}{i}-hole  (and the comparison
regions) lack  stellar populations  younger than $\sim400$  Myr, which
are otherwise abundant in the inner body of the galaxy.
We  conclude that  there is  no  convincing evidence  that the  SagDIG
\ion{H}{i}-hole  is   the  result   of  stellar  feedback,   and  that
gravitational and thermal instabilities in the gas are the most likely
mechanism for its formation.
}

%\textcolor{red}{red}


\maketitle

\section{Introduction}

Dwarf irregular  galaxies (dIrr)  have a crucial  role in  the ongoing
effort to understand galaxy  formation and evolution. Their relatively
simple structure and isolated location facilitate the study of stellar
evolutionary  phases;  in   particular  those  related  to  \emph{star
  formation} processes (\citealt{Mateo1998e}). Moreover, the interplay
between star formation and  the surrounding inter-stellar medium (ISM)
is  most  evident  in gas-rich  dIrr.  The  dIrr  of the  Local  Group
($D\lesssim1.1$ Mpc) have a particular place in these studies, as they
offer  the unique  possibility of  investigating  the \emph{interplay}
between  the resolved stellar  populations (whose  average metallicity
and  age can  be determined),  the surrounding  ISM, and  star forming
regions. The Sagittarius dwarf  irregular galaxy (also known as SagDIG
or  UKS  1927-177)  is  an  excellent  prototype  of  a  star-forming,
gas-rich, low-metallicity galaxy, where such studies can be conducted.

Cook (\citeyear{Cook1987}; hereafter C87) provided the first CCD study
of the  resolved stellar populations of  SagDIG, where the  use of two
intermediate passbands  allowed him to disentangle  between carbon and
M-type   stars.  Later   studies   by  \citet{Karachentsev1999j}   and
\citet{Lee2000c}  refined  the   distance  modulus  estimate  and,  at
($m-M$)$_{\circ}=$$25.13\pm0.25$,  SagDIG was  finally confirmed  as a
Local Group member. Interestingly, both studies agreed on SagDIG being
the \emph{most }metal-poor  ({[}Fe/H{]}$\sim-2.8$) star forming galaxy
in  the Local  Group. \citet{Momany2002}  advised that  a differential
reddening  scenario (where young  centrally-concentrated main-sequence
stars suffer higher reddening with respect to the older and off-centre
red  giants)  would  revise  the  SagDIG  photometric  metallicity  to
{[}Fe/H{]}$=-2.1\pm0.2$, thus  placing it within the  general trend of
the metallicity-luminosity relation for dIrr. This value was confirmed
by   \citet{Saviane2002}   who   derived   an  oxygen   abundance   of
$12+\log$(O/H)=7.26 for the brightest \ion{H}{ii} region in SagDIG.

A glimpse of the star formation  history of SagDIG was revealed by the
deep      Hubble     Space      Telescope      (HST)     study      by
\citet{Momany2005a}.       Color-magnitude      diagrams      reaching
$m_{F606W}\simeq27.5$  showed  the  presence  of a  conspicuous  young
($\lesssim1$ Gyr) population of main-sequence and He-burning blue-loop
stars. The identification  of the red clump indicated  the presence of
intermediate  ($1-10$  Gyr)  stellar  populations,  while  that  of  a
\emph{small} but genuinely-old ($\gtrsim10$ Gyr) red horizontal branch
stars  proved that  SagDIG, just  like any  other dwarf  galaxy, first
started  forming  stars a  Hubble  time ago.  \citet{Gullieuszik2008a}
conducted the  first near-infrared study of  SagDIG, and characterized
the evolved asymptotic giants and red supergiant populations.

On the other hand, the  spatial distribution of \ion{H}{i} in gas-rich
dwarf  galaxies  is  usually  clumpy,  and  single  \ion{H}{i}  clouds
($50-100$  pc) are  found near  (but do  not typically  coincide) with
regions   of   active  star   formation   and  \ion{H}{ii}   complexes
(\citealt{Hodge1994i};\citealt{Young1997j}; \citealt{hunter}).
On  a larger  scale,  the \ion{H}{i}  observations  of gas-rich  dwarf
galaxies reveal a  wealth of structure in the ISM;  some, but not all,
have       expanding      shells       (e.g.,      \ion{Holmberg}{ii};
\citealt{Rhode1999e}).
Cavities in the ISM of  gas-rich dwarfs are believed to originate from
the combined effects of photo-ionization, stellar winds and supernovae
explosions from  the sequential formation of massive  stars. There are
factors that contribute to making cavities a long-lived feature in the
ISM, namely (1)  the slow solid-body rotation of the  gas; (2) the low
gas densities;  (3) the reduced shear;  and (4) the  absence of spiral
density waves (\citealt{Stewart2000bd}).

Young  \&  Lo (\citeyear{Young1997j})  presented  high resolution  VLA
observations  showing  that  the  SagDIG \ion{H}{i}  content  extended
significantly  farther   out  than  the   stellar  component.  Without
convincing  signs  of  rotation,  the  \ion{H}{i}  gas  seemed  to  be
dominated by random motions. The SagDIG \ion{H}{i} content was divided
into  \emph{broad} ($\sigma=10$  km/s)  and \emph{narrow}  ($\sigma=5$
km/s)  components.   The   later  is  found  in  the   form  of  small
($8\times10^{5}$  M$_{\odot}$) clumps  that are  mainly in  the galaxy
centre, whereas the broad  component showed a regular distribution all
over the galaxy face.  This two-component structure was interpreted as
the  analogue  of  the  cold/warm  phase  structure  of  the  Galactic
\ion{H}{i}. Young \& Lo (1997)  estimated the total \ion{H}{i} mass of
SagDIG   to  be   $\sim9.3\times10^{6}$  M$_{\odot}$,   going   up  to
$\sim1.3\times10^{7}$  M$_{\odot}$ when  including  He.  Thus,  SagDIG
seems  to have maintained  a large  reservoir in  the form  of neutral
gas.  The  \ion{H}{i}  is  distributed  in an  almost  symmetric  ring
surrounding  a cavity  (\emph{horse-shoe} structure)  that  is, again,
attributed to the combined effects of stellar winds and supernovae.

All  together,   the  general  properties  of   the  resolved  stellar
populations,  \ion{H}{i}  and  \ion{H}{ii}  components of  SagDIG  are
known.  Nevertheless,  SagDIG is  projected  relatively  close to  the
Galactic    centre    ($\ell=21.06^{\circ},$$\,\,   b=-16.29^{\circ}$)
implying a \emph{strong} foreground Galactic contamination, attributed
to  the Galactic  thin  and thick  disk,  the halo,  and possibly  the
outskirts  of the  Galactic bulge.  This  is especially  true for  the
\emph{red   }stellar  populations   of   SagDIG  {[}\emph{young}   red
supergiants   (RSG),   \emph{intermediate-age}   carbon   stars,   and
\emph{old} asymptotic giant branch (AGB){]} which are heavily confused
by   the  Galactic  foreground   contamination,  preventing   (i)  the
reconstruction of  the SagDIG  star formation and  chemical enrichment
history; and  (ii) the  interplay between the  \ion{H}{i} distribution
and the SagDIG  stellar populations (in particular the  red ones). For
the above reasons HST \emph{second-epoch} data were obtained, enabling
a  differential proper motion  analysis of  the SagDIG  and foreground
field  populations.  In this  paper  we  present  the results  of  the
proper-motion decontamination process of  our 2005 HST SagDIG catalog,
which  is  made  available  in  its  entirety  via  the  link  to  the
machine-readable version.

\begin{figure*}
\begin{centering}
\includegraphics[width=12cm,height=14cm]{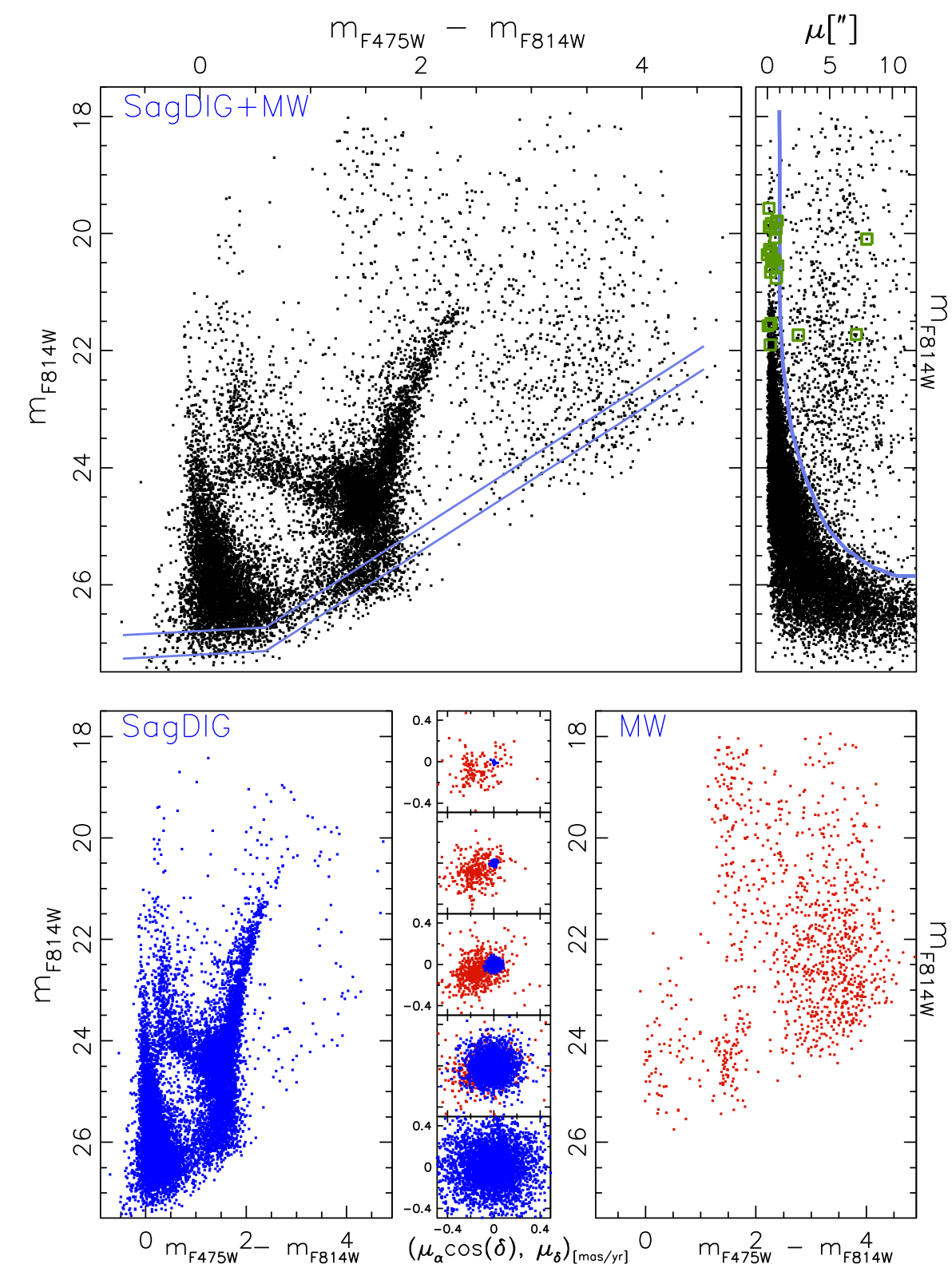}
\par\end{centering}

\caption{Upper left panel displays the \citet{Momany2005a} color-magnitude
diagram along with the $80\%$ and $50\%$ photometric incompleteness
levels. Upper right panel displays the $\ensuremath{\mu^{\prime\prime}}$
(see text) as a function of $m_{F814W}$, where open squares highlight
the position of the identified carbon stars and the heavy blue line
marks the separation between SagDIG and foreground stellar
populations.
Lower panels display the resultant proper-motion separated diagrams
of SagDIG and the Galactic foreground populations. Lower middle panel
displays the $5$ vector-point diagrams, in bins of $2.0$ magnitudes
per sub-panel, starting with $\ensuremath{m_{F606W}=18.0.}$ SagDIG
selected members are highlighted as blue points whereas Galactic foreground
is plotted in red. \label{fig: CMD-Selection-1}}
\end{figure*}

\section{Observations \& Data Reduction}

The first-epoch data set (GO$-9820$  P.I. Momany) consisted of $3$ HST
orbits;  dedicating one  orbit to  each  of the  $F475W$, $F606W$  and
$F814W$  filters.  Five  dithered  exposures  of  $410$  seconds  were
obtained  for each  filter.  The observations  were concluded  $2003$,
August  $18^{th}$. The  programmed second-epoch  data  set (GO$-10472$
P.I. Momany) consists of a  single HST orbit dedicated to observations
in  the $F814W$ filter,  in order  to avoid  possible filter-dependent
systematic errors.  The second-epoch data  were also divided  into $5$
exposures of $410$ seconds.  These observations were concluded $2005$,
June $21^{st}$. The resultant epoch separation is $\simeq1.85$ years.

Following the general recipes  given in \citet{Anderson2000h}, and for
deriving  precise   astrometric  measurements  and   a  more  accurate
assessment of  the errors,  we took particular  care in  designing the
\emph{dithering} pattern  of our second-epoch images in  terms of both
\emph{integral} and \emph{fractional}-pixel  offsets. We then measured
the positions and fluxes for  all detected stars in every $F814W\_FLT$
exposure.    For    this    task    we    used    library    effective
Point-Spread-Functions  and the software  programs, all  documented in
\citet{Anderson2006ag}.  We  then generated a  master list of  all the
stars,  and collated  all the  observations for  a given  star.  As in
\citet{Bedin2003b}, we used  the best distortion corrections available
(\citealt{Anderson2002}  and \citealt{Anderson2006ah}) to  correct the
raw positions measured from the $\_FLT$ exposures.

At  first,   we  identified   a  group  of   SagDIG  members   in  the
color-magnitude  diagram and  these were  adopted  as \emph{reference}
stars.  The coordinate  transformations from  a given  image  into the
reference frame  were derived using only  these \emph{reference} stars
(see  \citealt{Bedin2003b}  for details).  We  thus  ensured that  the
proper motion of all measured stars  is relative to the bulk motion of
the SagDIG members. This means  that the relative motion of the SagDIG
stars  should, to  within the  measurement errors,  be  centred around
zero. We then iteratively removed from the member list some stars that
had field-star-type motions, even though their colors placed them near
the fiducial SagDIG sequences.

Lastly, in order to minimize the influence of uncorrected distortion
residuals on the transformations into the reference frame, \emph{local
transformations} based on the nearest $\sim50$ well-measured SagDIG
stars were used for \emph{all} detected stars. Taking all these precautions
into account, the position of stars with $>500$ DN in their brightest
pixel (in a given single image) proved to posses an error of $<0.05\%$
of a pixel. We note that the relatively high background level in our
single images ($\sim65$ DN), and the relative \emph{youth} of the
ACS/WFC detectors (at 2005), resulted in a negligible astrometric
effect caused by charge-transfer efficiency problems.

\section{Decontaminated Catalog }

The results of this differential proper motion analysis are summarized
in Fig.\,\ref{fig: CMD-Selection-1}. The upper left panel displays the
original    Momany   et   al.    \citeyearpar{Momany2005a}   composite
$\ensuremath{m_{F814W},}$          ($\ensuremath{m_{F475W}-m_{F814W}}$)
color-magnitude  diagram  towards   SagDIG.  The  Galactic  foreground
contamination  is   visible  as   two  broad  vertical   sequences  at
($\ensuremath{m_{F475W}-m_{F814W}}$)$\ensuremath{\simeq1.5}$        and
$\simeq3.0$, in between  which the SagDIG red giant  branch, red clump
and  red  supergiant  population  are  recognizable  thanks  to  their
\emph{relatively} higher stellar density. The young stellar population
of SagDIG  (composed of  main sequence and  blue supergiant  stars) is
mostly                                                             blue
${[}(\ensuremath{m_{F475W}-m_{F814W}})\ensuremath{\lesssim1.0}{]}$
thus  suffering  negligible Galactic  contamination.  The upper  right
panel  plots   $\ensuremath{\mu^{\prime\prime}}$  as  a   function  of
$\ensuremath{m_{F814W}}$  magnitude,  which is  the  filter in  common
between        the       data        sets        of       the        2
epochs.    $\ensuremath{\mu^{\prime\prime}}$     is    expressed    in
milliarcseconds          and         is          calculated         as
$\ensuremath{\sqrt{dx^{2}+dy^{2}}}$  (where  $dx$  and  $dy$  are  the
derived offsets in pixels between  the 2 epochs) normalized to the ACS
pixel-scale              and              epoch             separation
${[}i.e.\ensuremath{\frac{\textrm{(}\sqrt{dx^{2}+dy^{2}}~\times~0\farcs05)}{1.85}\times1000}{]}$.
The   distribution,   especially    for   magnitudes   brighter   than
$\ensuremath{m_{F814W}\sim24.0}$ shows a  clear separation between the
SagDIG             stellar             populations             (having
$\ensuremath{\mu^{\prime\prime}}\ensuremath{\lesssim1.0}$)  and  those
belonging     to    the     Milky    Way,     showing     a    broader
$\ensuremath{\mu^{\prime\prime}\gtrsim2.0}$.    In   particular,   and
highlighted  as open  squares, the  sample of  Carbon stars  (see next
Section) shows a  clear separation between SagDIG members  and a group
of 3 Galactic carbon stars.

For    $\ensuremath{m_{F814W}\gtrsim24.0}$    increasing   astrometric
uncertainties do  not allow a clear-cut separation  between the SagDIG
and   Milky  Way  populations.   Following  various   experiments,  we
conservatively  draw a separating  line that  maximizes the  number of
SagDIG  members. A  closer look  at the  upper panels  shows  that for
$\ensuremath{m_{F814W}\gtrsim24.0}$                                 and
$2.5\lesssim(\ensuremath{m_{F475W}-m_{F814W}})\ensuremath{\lesssim4.0}$
the star counts of these Galactic foreground dwarfs (mostly due to the
thick disk  population with little  contribution by the thin  disk and
the Galactic halo populations) drop rapidly. This is solely due to the
photometric incompleteness of our first-epoch data. The thick lines in
the  upper left  diagram display  the $80\%$  and  $50\%$ completeness
levels as  derived from the artificial star  experiments, presented in
\citet{Momany2005a}.

The  upper panels  display the  {\it raw}  diagrams as  a  function of
$\ensuremath{m_{F814W}}$  which was used  for the  differential proper
motion analysis. The lower  panels display the resultant {\it cleaned}
diagrams, also as a function of $\ensuremath{m_{F814W}}$.
The  lower  middle  panel  displays   an  enlarged  view  on  the  $5$
vector-point  diagrams, where  each panel  shows the  distribution for
bins of  $2.0$ magnitudes starting  $\ensuremath{m_{F814W}=18.0.}$ The
vector-point  diagrams  thus  include  the SagDIG  (blue  points)  and
Galactic  (red points) populations.  The lower  left and  right panels
display the  proper-motion decontaminated  diagrams of the  SagDIG and
Milky Way population, respectively.

\section{The SagDIG AGB Stellar Populations}

Prior to our proper-motion decontamination, the disentangling
of the SagDIG red stellar populations (being RSG, oxygen and carbon-rich
AGB) was hindered by the foreground Galactic contamination. To a lesser
extent, and thanks to their higher number density, RGB stars allowed
earlier determination of the luminosity of the tip of the giant branch. In
this section we address the properties of these unveiled populations,
confident that the decontamination process allows a reliable membership
down to at least $F814W\simeq24.0$ (i.e. the red clump level).

Being brighter than the tip of the red giant branch, AGB stars
are the first population to be detected in distant galaxies, and they have
 long been used (e.g. \citealt{Letarte2002}) to investigate
the distribution of intermediate-age stellar populations. The Thermally
Pulsating AGB evolutionary phase is characterized by mixing mechanisms
that draw inner processed material to the outer stellar atmospheres.
This alters the primary oxygen-rich composition and gives rise to
carbon-rich stars (where the ratio of $C/O$ atoms is $>1$). Iben
\& Renzini \citeyearpar{Iben1983c} were the first to show that the
conversion from $O-$rich to $C-$rich is facilitated at lower metallicities.
Indeed, in metal-poor stars, with low $O$ abundance, less thermal
pulses are required to produce $C/O>1$. Overall, the number ratio
of $C-$rich to $O-$rich stars allows one to infer the metallicity of
a stellar population of a given age (\citealt{Battinelli2005}). However,
the $C/M$ ratio is extremely sensitive to the star formation history, 
since the specific production of $C-$and
$O-$rich stars is strongly dependent on the initial stellar mass.
This is particularly true for intermediate-mass stars for which the
production of TP-AGB have a well-pronounced maximum (see \citealt{Girardi2013c}).
Intermediate-mass stars are predominately $C-$rich and therefore,
at a given metallicity, the $C/M$ ratio is expected to be higher
in stellar systems with intermediate-age ($\sim1-2$ Gyr) stellar
populations. Consequently, this ``degeneracy'' has to be accounted
for when the $C/M$ ratio is used to probe the \emph{metallicity}
of stellar populations (\citealt{Sibbons2012}). 

\begin{figure}
\begin{centering}
\includegraphics[scale=0.45]{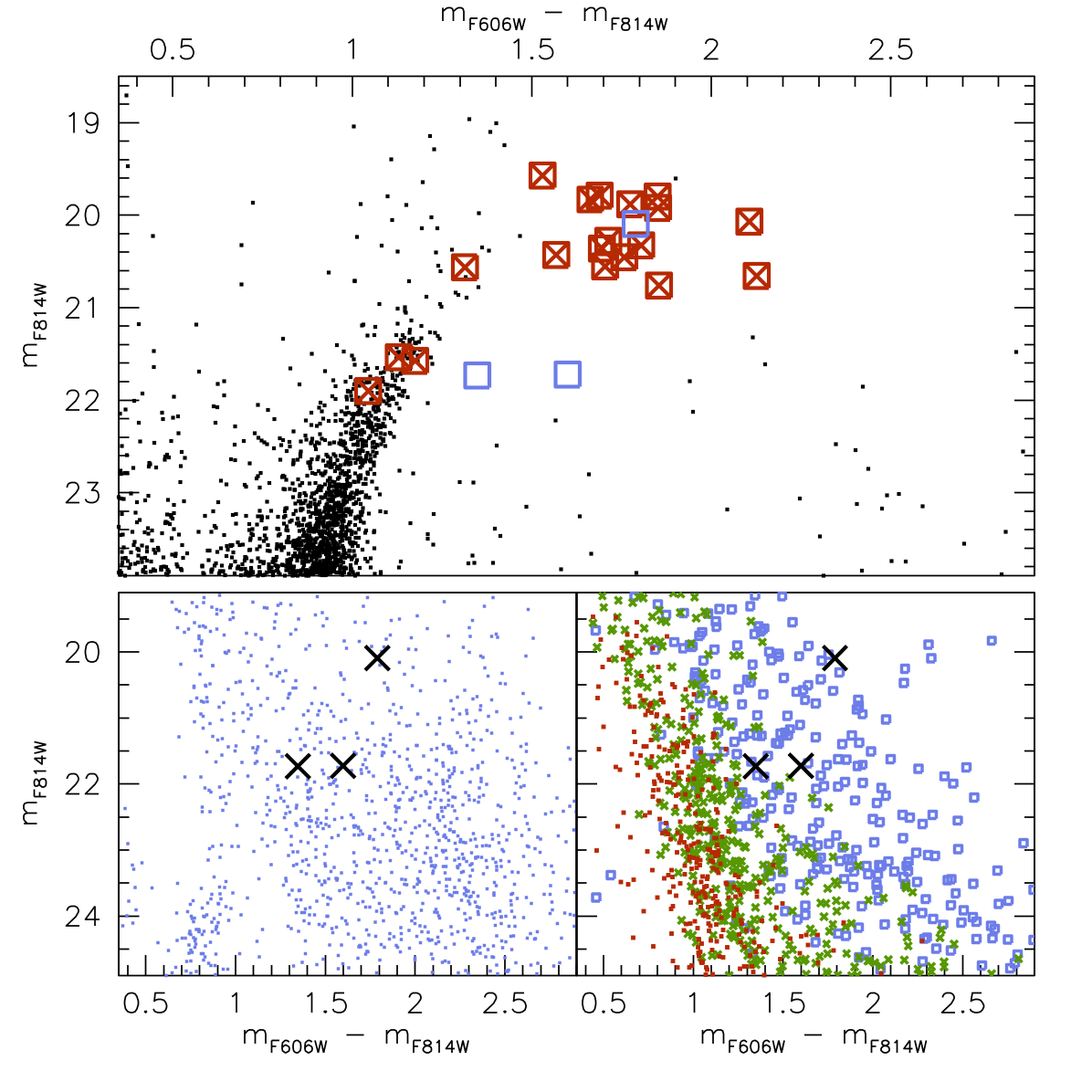}
\par\end{centering}
\caption{Upper panel displays the SagDIG carbon stars highlighted on the proper-motion
decontaminated $m_{F814W},(m_{F606W}-m_{F814W})$ color-magnitude
diagram. Members are displayed as squared-crosses, while those of
Galactic origin are shown as open square symbols. Lower left panel
displays the Galactic foreground proper-motion decontaminated diagram,
where the $3$ crosses highlight the carbon stars with Galactic membership.
Lower right panel shows a Trilegal simulation of the Galactic field
along the SagDIG line of sight (thin/thick disk and Galactic halo are
plotted as open squares, crosses and dots, respectively). \label{fig:Carbon_CMD}}
\end{figure}

One of the most efficient methods to disentangle and classify $C-$
and $O-$rich stars is based on the narrow-band photometry technique
pioneered by Cook et al. \citeyearpar{Cook1986ah}. The method
relies on the use of a $CN$ and $TiO$ filters that are respectively
centred on molecular bands present in the spectra of $C-$rich and
$O-$rich stars. However, although the $CN-TiO$ color index is a
powerful $C-$ and $O-$rich separator, this technique cannot disentangle
dwarf from giant stars. Consequently, when applied to select the AGB
populations in nearby galaxies, control fields are used to estimate
the Milky Way foreground contamination.

\subsection{Carbon-rich Stars}

Cook \citeyearpar{Cook1987} identified a sample of $26$ $C-$stars
in SagDIG. \citet{Demers2002b} presented another sample of $C-$stars,
$8$ of which were in common with the Cook sample. Of both samples
$21$ $C-$stars were identified within our, relatively smaller, ACS@HST,
field of view. Table\ref{tab:C-stars-1} presents the photometric
properties of these stars along with coordinates and proper-motion
displacement. Of particular interest is the identification of 3 $C-$stars
whose displacement is consistent with a Galactic origin (c.f. upper
right panel of Fig.\,\ref{fig: CMD-Selection-1}). 

The upper panel of Fig.\,\ref{fig:Carbon_CMD} shows the distribution
of the $21$ $C-$stars in the decontaminated color-magnitude diagram.
The majority of the SagDIG $C-$stars are around $\simeq1$ magnitude
brighter than the tip of the red giant branch, and display redder colors.
This is consistent with the $C-$star identification in near-infrared
diagrams \citep{Nikolaev2000g} having $(J-K)\gtrsim1.4$, and confirmed
in the near-infrared, SagDIG study of \citet{Gullieuszik2008a}. \citet{Battinelli2005a}
addressed the \emph{standard candle} aspect of carbon stars and concluded
that their average $M_{I}$ luminosity is viable as long as the parent
galaxy contains some hundred $C-$stars or more. Indeed, their Fig.\,3
shows that for large galaxies within $-20\lesssim M_{V}\lesssim-15$
the average $M_{I}$ carbon stars luminosity is basically constant
at around $-4.6\pm0.1$. Instead, smaller galaxies (i.e. $M_{V}\gtrsim-15$)
show a larger dispersion. In the case of SagDIG (for which we derive
a $M_{V}=-11.0$) the average $M_{I}$ of the $16$ carbon stars,
that are redder than the RGB tip, is $M_{I}\simeq-4.95$, in excellent
agreement with the \citet{Battinelli2005a} results. Our value is
the result of transforming the $m_{F814W}$ magnitudes to $I-$Johnson
Cousins systems and applying the reddening of $E_{(B-V)}=0.12$ and
$(m-M)_{\circ}=25.10$ (as derived in \citealt{Momany2005a}). 

The upper panel of Fig.\,\ref{fig:Carbon_CMD} also shows a
group of $3$ confirmed $C-$stars clustering on the SagDIG red giant
branch, just below the tip level. The identification of such faint,
blue $C-$stars can be evidence of binary evolution, where the red
giant has accumulated carbon-processed material from the primary component
of the system, now an invisible and cooling white dwarf. In this regard,
the faint and blue $C-$star giants simply reflect $C-$polluted
atmospheres that were produced externally, as opposed to the AGB $C-$stars
whose carbon enhancement is intrinsic to the star itself. 
Stellar evolutionary models can actually produce TP-AGB
$C-$stars below the red giant branch tip level (\citealt{Marigo2013b}).
The very identification of such faint, blue $C-$stars would therefore
indicate that the metallicity of the intermediate-age stellar
populations of SagDIG is very low. This is not surprising, indeed \citet{Saviane2002}
derive a metallicity of $12+\, log(O/H)=7.26$ for the one
and only \ion{H}{ii} region in SagDIG.

The lower 2 panels of Fig.\,\ref{fig:Carbon_CMD} address the possible
origin of the $3$ Galactic $C-$stars. The left panel displays the
Milky Way color-magnitude diagram as derived from the proper-motion
decontamination process. We compare this \emph{observed} diagram to
a \emph{simulated} diagram according to the Trilegal Milky Way model
(\citealt{Girardi2012c}) that includes thin and thick disk and halo
components, all simulated in the HST/ACS photometric system. Relying
solely on their position in the diagrams, the $3$ $C-$stars are
consistent with being Milky Way thin disk \emph{dwarf} stars. This
is particularly interesting given the results from the Sloan Digital
Sky Survey (\citealt{Downes2004b}) and the Digitized First Byurakan
Survey (\citealt{Gigoyan2012c}) addressing the detection of high-latitude
carbon stars. As reported in these studies, faint carbon stars were
initially thought to be distant red giant branch stars. However, at
least $50\%$ of the faint carbon stars detected by the Sloan survey
displayed parallaxes and/or high proper-motions that are more appropriate
for \emph{nearby}, main-sequence stars. The relatively high proper-motion
of our $3$ $C-$stars added to their position in the color-magnitude
diagram all point to the \emph{first} reliable detection of \emph{low-latitude}
Galactic carbon dwarfs, most probably belonging to the thin disk population. 

\begin{figure}
\begin{centering}
\includegraphics[scale=0.45]{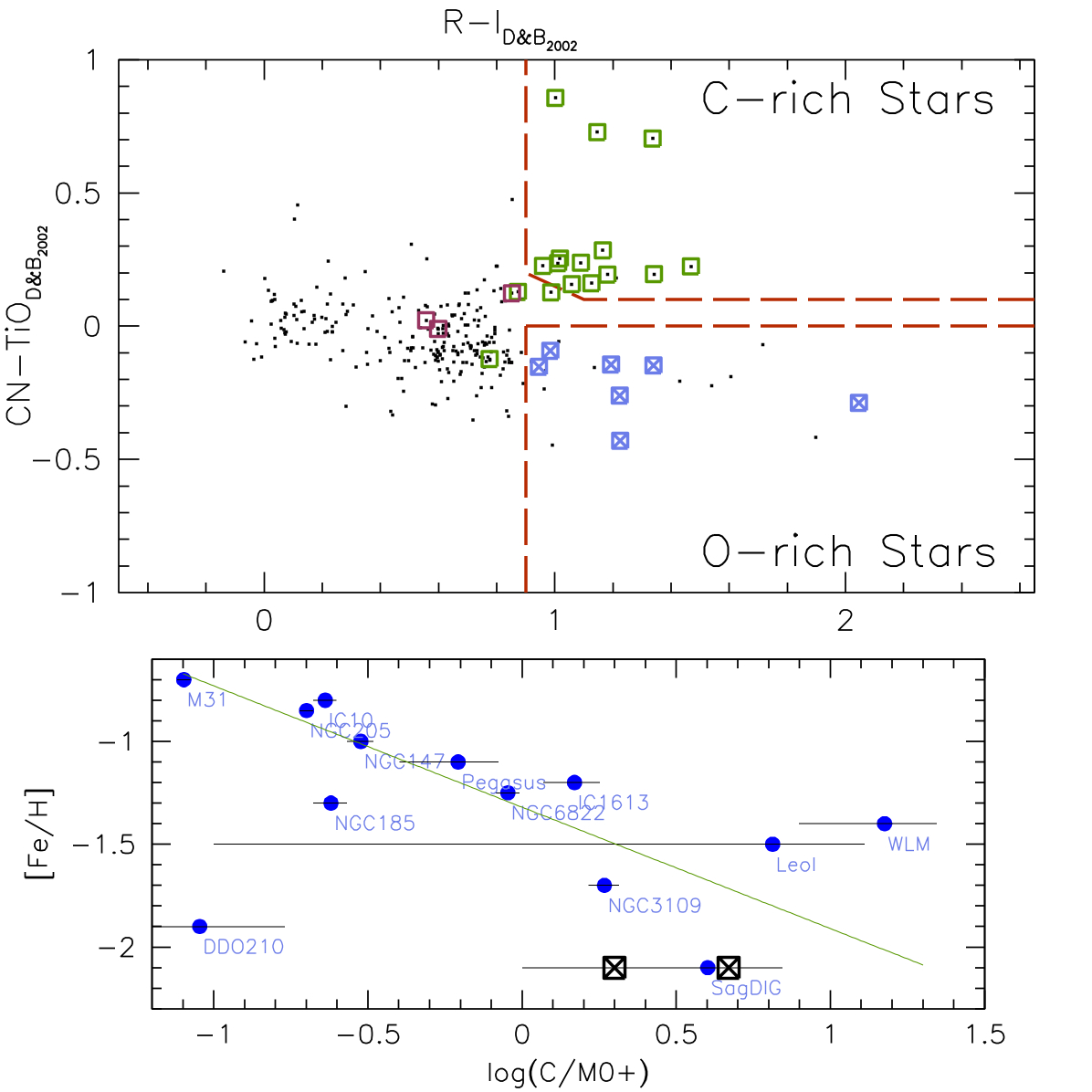}
\par\end{centering}

\centering{}\caption{Upper panel displays the $(CN-TiO),$ $(R-I)$ two-color diagram,
with highlighted the \citet{Demers2002b} selection boxes for carbon
and oxygen-rich stars. Dark-red open squares are carbon-stars (from
\citealt{Cook1987}) falling below the red giant branch tip. Lower
panel displays a reproduction of the {[}$Fe/H${]}, $log(C/M0+)$
diagram from Battinelli \& Demers (\citeyear{Battinelli2005a}), with
the new $C/M0+$ determination for SagDIG following the proper-motion
decontamination process. \label{fig:C/M ratio}}
\end{figure}

\subsection{Carbon/Oxygen Ratio\label{sub:Carbon/Oxygen-Ratio}}

In this  paper we adopt  the $C/M$ star selection  criterion following
Battinelli \& Demers (\citeyear{Battinelli2005a}), which refers to the
selection of stars of spectral  type $M0$ and later, denoted as $M0+.$
The  $(CN-TiO),$  $(R-I)$  two-color   diagram  (see  upper  panel  of
Fig.\,\ref{fig:C/M   ratio})  guarantees,  \emph{in   principle},  the
identification  of $C-$stars that  are \emph{certain}  galaxy members.
Indeed, the  major difficulty with  the derivation of the  $C/M$ ratio
resides  in the  selection  of  the oxygen-rich  star  sample that  is
foreground-contamination    \emph{free}.     To    this    end,    the
foreground-contamination  is usually  estimated  using control  fields
around  the galaxy  of  interest. In  this  regard, our  proper-motion
decontamination process  provides the first  independent evaluation of
the $C/M0+$ derivation process. \citet{Demers2002b} kindly provided us
with  their SagDIG  $4-$filter catalog  and  this was  matched to  our
SagDIG member catalog, the results of which are presented in the upper
panel  of   Fig.\,\ref{fig:C/M  ratio}.  The   identification  of  the
$C-$stars is  straightforward and shows  the presence of a  few $blue$
$C-$stars\footnote{We note that the $blue$ $C-$stars \emph{all} belong
  to  the \citet{Cook1987}  sample.} falling  outside  the carbon-star
selection box,  and includes the  $3$ carbon-stars falling on  the red
giant branch (dark-red open squares).   To ensure the selection of AGB
oxygen-stars,   Battinelli   \&  Demers   (\citeyear{Battinelli2005a})
introduce  another   selection  limit,   that  is,  the   adoption  of
$M_{bol}=-3.5$  for  the  AGB  star bolometric  magnitudes.  We  allow
selection down to $\sim0.5$ magnitude  below the red giant branch tip,
and overall  count $7$  oxygen-rich stars, \emph{vs}  $14$ carbon-rich
stars.

\begin{figure}
\begin{centering}
\includegraphics[scale=0.45]{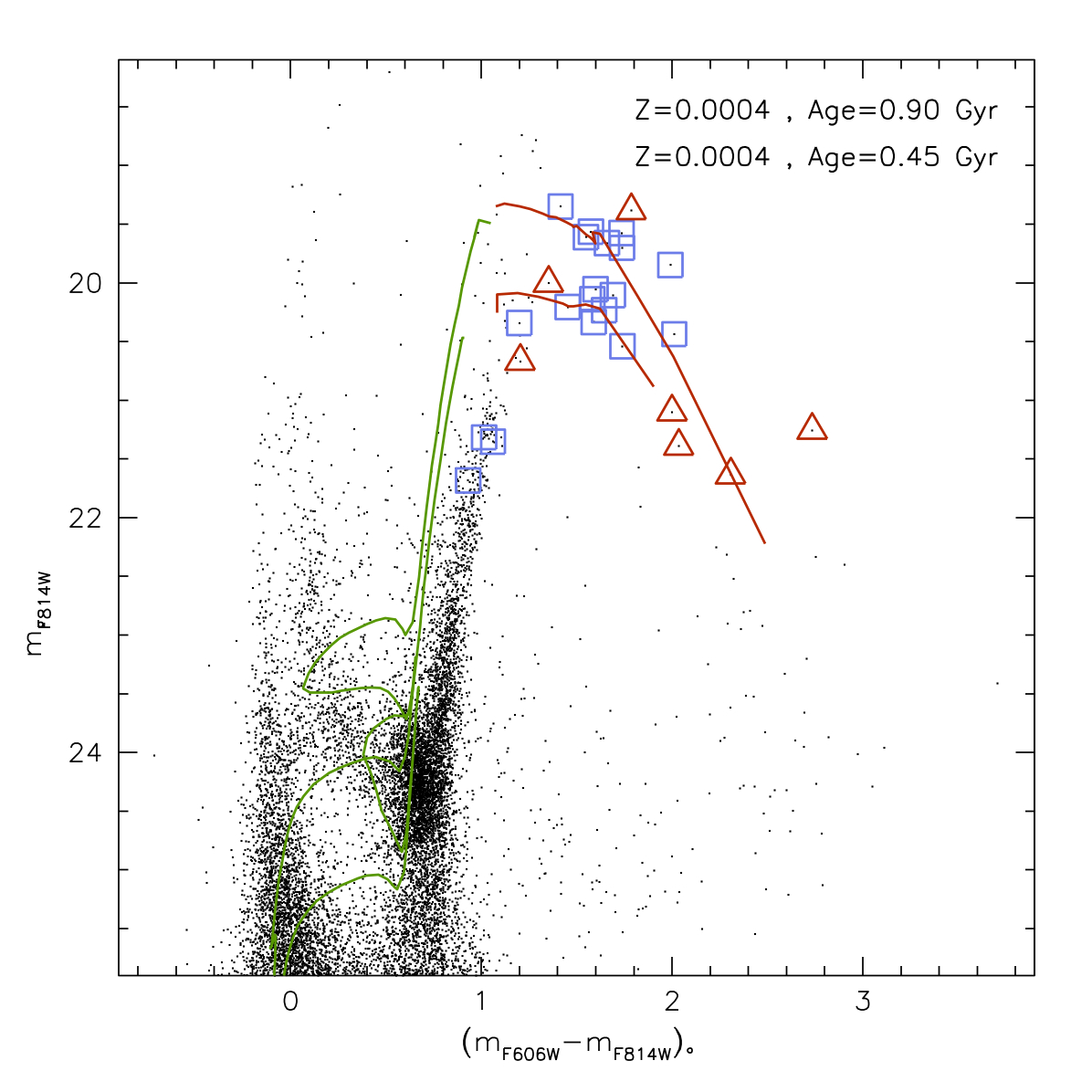}
\par\end{centering}
\centering{}\caption{The dereddened $m_{F814W},(m_{F606W}-m_{F814W})$ SagDIG color-magnitude
diagram highlighting the carbon stars (open squares) and $2$ sets
of isochrones from the \citet{Marigo2008a} library. Open triangles
are stars defined as oxygen-rich (see Sec.\ref{sub:Carbon/Oxygen-Ratio}).\label{fig:isochrones_Cstars}}
\end{figure}

The identification of the $7$$ $ oxygen-rich stars (open triangles
in Fig.\,\ref{fig:isochrones_Cstars}) warrants some attention. Indeed,
$4$ of these are very \emph{red} ($m_{F606W}-m_{F814W}\gtrsim1.7$)
and \emph{faint} (with luminosities below the red giant branch tip
level; $m_{F814W}=21.3$). Interestingly, the $4$ \emph{red/faint}
$O-$rich stars appear to lie on the \emph{redder/fainter} elongation
of the carbon star isochrones. This hints at the possibility that
these are $O$-rich, dust-enshrouded stars with high bolometric luminosity,
and therefore one would expect them to be relatively \emph{bright}
at infrared wavelengths. To explore this possibility, we re-examined
the SagDIG near-IR photometry from \citet{Gullieuszik2007c}. The
$4$ \emph{red/faint} $O-$rich stars have $K_{s}$ magnitudes
that are \emph{fainter} than the SagDIG RGB tip (at $K_{s}\sim18.5$),
and $J-K_{s}$ colors that are typical of normal RGB stars. Furthermore,
no \emph{AllWISE }(Cutri et al. \citeyear{Cutri2014}) counterparts
were found within $1^{\prime\prime}$ of the centre of the \emph{$4$
$O-$}rich stars. With the lack of strong emission at near ($2.2\,\mu$m)
and mid-infrared ($3.4,$ $4.6,$ $12,$ and $24$\,$\mu$m) wavelengths,
one is led to doubt that the $4$ \emph{red/faint} $O-$rich candidate
stars are \emph{genuine} oxygen-rich stars.

On the other hand, they may have a genuine oxygen-rich origin if we postulate 
that these 4 stars \emph{need not to be} \emph{massive}.
This \emph{possibility} can be easily accommodated granted the fact
that these $O-$rich candidate stars reside in a \emph{significantly}
metal-poor environment, a constraint that is easily met by the \emph{very}
low metallicity of SagDIG ({[}Fe/H{]}$=-2.1$ or $Z=0.00025$). Indeed,
the analysis presented in \citet{Nanni2013a} (describing the $ $
$C\leftrightarrows O$ rich evolution during the thermal pulsating
AGB) provides a framework within which the $4$ \emph{red/faint} stars
are conceived as genuine $O-$rich stars. In particular, their \emph{case-d}
scenario (for a relatively massive $4$ M$_{\odot}$ and $Z=0.001$
model star) shows how the transition from $C-$rich to $O-$rich takes
place following the ignition of the Hot-Bottom-Burning process, where
the efficiency of the CN cycle (depleting carbon) determines the transition
to the $O-$rich spectral type. 

Extrapolating the \citet{Nanni2013a} \emph{case-d} ($4$ M$_{\odot}$,
$Z=0.001$ and $0.15$ Gyr) scenario for the \emph{very} low metallicity
of SagDIG, would allow for the appearance of $\sim2-2.5$ M$_{\odot}$,
$Z=0.00025,$ and $0.45-0.90$ Gyr faint $O-$rich stars, along the
extension of the $C-$rich stars (as seen in Fig.\,\ref{fig:isochrones_Cstars}).
Similarly, as also described in their \emph{case-e, }the alternating
effects of the third dredge-up (enhancing the production of carbon-rich
stars) and Hot-Bottom-Burning (enhancing the oxygen-rich stars) would
reflect in \emph{multiple transitions} across $C/O=1$, and $O-$rich
stars can materialize along the extension of the \emph{red/faint}
$C-$rich tail. One last alternative scenario (permitting the
presence of the $4$ \emph{red/faint} $O-$rich stars) concerns the
possibility that these are less-massive ($\lesssim2$ M$_{\odot}$),
older ($\gtrsim1$ Gyr) AGB stars possessing intrinsically reddened
$J-K_{s}$ colors. In conclusion, the origin of these $4$ \emph{red/faint}
oxygen-rich candidate stars remains uncertain, making them ideal targets
for a spectroscopic follow-up.

The lower panel of Fig.\,\ref{fig:C/M ratio} is a re-production
of the Battinelli \& Demers (\citeyear{Battinelli2005a}) {[}$Fe/H${]}--$log(C/M0+)$
anti-correlation for a sample of Local Group dwarf galaxies, where
the straight line displays their least-square linear fit to the data.
The squared-crosses delimit the newly derived SagDIG $C/M0+$ ratio
including, or not, the $4$ $O-$rich \emph{red/faint}
stars. Overall, both values fall within the expected error of Battinelli
\& Demers (\citeyear{Battinelli2005a}), emphasizing the important
role played by SagDIG (being the most-metal poor galaxy among the
Local Group galaxies) in establishing the zero-point of the {[}$Fe/H${]}--$log(C/M0+)$
anti-correlation. 

\subsection{Isochrone Fitting of the Carbon Stars}

The thermally pulsating AGB phase is certainly the most poorly understood
evolutionary phase. Indeed, this phase includes complex processes
such as the third-dredge up, Hot Bottom Burning, mass-loss through
super winds and molecular opacity calculation that has defied reliable
inclusion of this phase in many of the available isochrone libraries.
The recent study by \citet{Marigo2008a} has not only provided a detailed
and improved treatment of the former mentioned processes, but also
calibrated their final isochrone set to known observable properties
of TP--AGB stars. In particular, their isochrones are offered in several
photometric systems. We therefore employ their HST/ACS set of isochrones
\emph{directly} on our data, and cautiously attempt the derivation
of the C-stars' age and metallicity.

Figure\,\ref{fig:isochrones_Cstars} displays a  set of isochrones from
the         \citet{Marigo2008a}        library         upon        the
$m_{F814W},(m_{F606W}-m_{F814W})$    color-magnitude   diagram,   with
$C-$stars highlighted as open  squares.  The bolometric corrections of
Aringer et  al. \citeyearpar{Aringer2009},  and a dust  composition of
$60\%$ Silicate $+40\%$ $AlOx$ for oxygen-rich stars, and $85\%$ $AMC$
and   $15\%$   $SiC$   are   applied.   Following   the   recipes   by
\citet{Sirianni2005d}   and    assuming   a   distance    modulus   of
$(m-M)_{\circ}=25.10$      and     $E_{(B-V)}=0.12$,      we     adopt
$A_{F475W}=0.437$,                $A_{F606W}=0.335$,               and
$A_{F814W}=0.221$.  Figure\,\ref{fig:isochrones_Cstars} summarizes our
effort to  derive the $C-$stars' age and  metallicity.  For relatively
metal-poor   $Z=0.0004$   isochrones   (reasonable  for   the   SagDIG
metallicity) we  find that $0.45$  and $0.90$ Gyr old  isochrones best
reproduce  the   colors  and  luminosities  of   the  $C-$stars.  This
particular point will  be further investigated (Held et  al. in prep.)
when reconstructing the star formation history of SagDIG.

\section{The \ion{H}{i} Hole}

In \citet{Momany2005a} we considered various possibilities
for the formation of the large \ion{H}{i}-hole in the South-West
of SagDIG. In particular, we estimated whether the integrated luminosity
of the stars within the hole was sufficient to represent the remnant
population of {\emph{a ``cluster''}}  whose massive
stars created the hole via supernovae explosions and winds
(i.e. stellar feedback). The conclusion was
that although the integrated light over the entire area of the hole
{\emph{was}}  sufficient to represent such a population,
the spatial distribution of stars within the hole shows no concentration
towards the centre of the \ion{H}{i}-hole, and rather just displays a 
gradual increase of light intensity towards the optical centre of
SagDIG.
Therefore,  no  evidence  was  found  of  ``any  age-specific  stellar
population   that    is   preferentially   distributed    within   the
\ion{H}{i}-hole''.  Unless supernovae  are approximately two orders of
magnitude  more effective at  imparting mechanical  energy to  the ISM
than  currently   thought  (so  that  the   putative  remnant  stellar
population in  the hole could go  undetected), the only  hope left for
the supernovae driven  wind hypothesis is that the  hole is actually a
very  {\emph{ ``ancient structure''}}  formed by  the supernovae  of a
``cluster''  of stars  that has  since faded  so much  that  they were
undetected by the HST data used in \citet{Momany2005a}.  The HST data,
now  cleaned  for  Galactic  foreground  contamination,  allow  us  to
re-investigate   this   (``ancient   structure'')   possibility   more
rigorously,   and  thereby   test  our   previous  \citet{Momany2005a}
conclusions (which were based  {\emph{solely}} on the integrated light
within the hole).

At first glance (c.f. Fig.\,\ref{fig: CMD-Selection-1})
one might assume that the proper-motion decontamination process adds
little to our previous examination, since the data set
remains the same. 
In  particular,   the  {\emph{massive}}  and   {\emph{young}}  stellar
populations of  SagDIG are relatively bluer than  the Galactic turnoff
color of {\emph{low-mass}} and {\emph{old}} main-sequence stars, which
(projected along the line of sight) form a loose and vertical sequence
around    $m_{F475W}-m_{F814W}\sim1.5$.   This    is    however   true
{\emph{only}}  for  the  SagDIG (Hydrogen-burning)  main-sequence  and
(Helium-burning)   blue-supergiant  stellar   populations   which  are
relatively robust  to Galactic foreground contamination.  On the other
hand,  the  {\emph{red-supergiant}}  and  AGB populations  (which  run
parallel to the main SagDIG  red giant branch) {\emph{are}} subject to
foreground contamination by {\emph{old }} Galactic main-sequence stars
(having      reached       or      approaching      their      turnoff
level/colour). Figure\,\ref{fig:isochrones_Cstars} further illustrates
this  point: examining  the $\sim0.45$  Gyr ($Z=0.0004$)  isochrone in
between  $20.5\le  m_{F814W}\le23.5$,   one  can  appreciate  how  the
red-supergiant   population   of   SagDIG   is  outnumbered   by   the
blue-supergiants.   In particular,  the red-supergiant  and  young AGB
population have  such low-frequency that  their numbers can  be easily
veiled even  by a low foreground contamination.  Most importantly, the
$\sim0.90$  Gyr isochrone  in  Fig.\,\ref{fig:isochrones_Cstars} is  a
reminder of how the identification  of a handful of red-supergiant and
young AGB  stars (proper-motion confirmed  members) would allow  us to
trace {\emph{older}}  stellar clusters,  within the hole.  Indeed, the
detection  of a  few red  supergiant/AGB stars  at $m_{F814W}\sim21.0$
would  trace the  presence of  a ``{\emph{stellar  cluster}}''  with a
turnoff level  at $m_{F814W}\sim26.3$ (at  the detection limit  of our
data).  Thus,  the proper-motion  decontamination  process provides  a
unique   and  un-biased   $\sim1\,$   Gyr  look-back   time  for   the
\ion{H}{i}-hole star formation history; tracing the remnant population
of  possible,  past and  defunct,  {\emph{clusters}}  that might  have
provided a number of supernovae.

To guide us in our new analysis, addressing the star formation
history within the \ion{H}{i}-hole, we refer to Fig.\,\ref{fig:HI-HOLE}
which displays the GALEX 
far-ultraviolet  image  of SagDIG  upon  which the  \citet{Young1997j}
\ion{H}{i} contours are superimposed. Besides highlighting the highest
density \ion{H}{i} clumps and void (\ion{H}{i}-hole), we also show the
ACS   footprint   which   emphasizes   the  full   coverage   of   the
\ion{H}{i}-hole        region.         The        SagDIG        centre
{[}(R.A.,Dec.)$=(292.499518,-17.680675)]$  was derived  as  the median
position of MS,  BSG and RSG stars  (of all ages) and is  shown with a
cross. To  allow a direct  comparison between the  stellar populations
present within the  \ion{H}{i}-hole and appropriate comparison fields,
we select $4$ elliptical regions  that are equidistant from the SagDIG
centre. The South-Western (SW) elliptical region (i.e. \ion{H}{i}-hole
centre)  will be  compared with  areas with  (i) ``{\emph{moderate}}''
\ion{H}{i}   content   (i.e.     North-Western   region)   and,   most
interestingly,  {\emph{(ii) high}}  \ion{H}{i} column  density regions
(North-Eastern and South-Eastern ellipses)
\footnote{The      only      known     \ion{H}{ii}      region
    (\citealt{Strobel1991})  is  at   the  rim  of  the  South-Eastern
    \ion{H}{i} clump.}.

The upper panels of Fig.\,\ref{fig:HI-HOLE-Simulations}
display the {\emph{observed and proper-motion decontaminated}} 
color magnitude diagram of the stellar populations present in the
selected SE and SW  (\ion{H}{i}-hole centre) equivalent regions. 
To minimize the poor statistics of the color-magnitude diagrams
in the SagDIG comparison fields, regions SE, NE,
and NW have been merged together in order to infer
a ``global'' star formation history of the comparison fields.
We derived the star formation history (SFH) of the \ion{H}{i}-hole and
comparison  fields by constructing  {\emph{synthetic}} color-magnitude
diagrams  (the lower panels)  using the  ZVAR simulator  (G. Bertelli,
priv.  comm.,  which in  turn  is based  on  the  uniform database  of
solar-scaled   evolutionary  tracks  of   \citealt{Girardi2000c})  and
comparing  these  to  the {\emph{observed}}  color-magnitude  diagrams
using  a maximum  likelihood  method. For  these  simulations we  have
adopted a  simple metal-enrichment law with  metallicity linearly (and
monotonically) increasing from {[}Fe/H{]}$=-2.1$ (as suggested for the
old stellar  populations by our  previous study \citealt{Momany2005a})
to {[}Fe/H{]}$=-1.67$  (assuming for the youngest  stars a metallicity
similar to that  of the \ion{H}{ii} region in  SagDIG {[}1/50 solar{]}
by \citealt{Saviane2002}). The present  simulations do not account for
internal differential  reddening (which probably affects  the color of
the  young   blue  plume  \citealt{Momany2002}),   assume  a  standard
power-law Initial  Mass Function  with $\alpha=-2.35$ in  the interval
between $0.1-100$ M$_{\odot}$, a binary fraction of $35\%$, and employ
the photometric  errors and incompleteness levels as  derived from the
artificial star simulations (see \citealt{Momany2005a}).

Although   the   proper-motion   decontaminated  SagDIG   catalog   is
photometrically complete  down to $m_{F606W}\simeq26.5$,  the selected
stellar  samples  in the  \ion{H}{i}-hole  and  comparison fields  are
basically    {\emph{void}}    of    young    stars    brighter    than
$m_{F606W}\simeq23.0$. We emphasize that this {\emph{``null-statistics
    regime''}} for stars brighter than the red clump level simply does
not  allow the reconstruction  of the  star-formation history  in this
time  interval.  In   particular,  stellar  populations  younger  than
$\sim350-400$ Myr can be {\em  ruled out} for both the \ion{H}{i}-hole
and the comparison fields (as evident from the over-plotted isochrones
in Fig.\,\ref{fig:HI-HOLE-Simulations}).
 The shortage of  young ($\le400$ Myr) Hydrogen and Helium-burning
  stellar populations  in the  selected elliptical regions  is however
  fully explained by their distance from the centre of SagDIG and star
  formation   sites,  which   hosts  abundant   MS  and   BSG  stellar
  populations.
This is clearly illustrated in the right-most panel of
Fig.~\ref{fig:HI-HOLE-Simulations} .
  A reconstruction of the \emph{recent } ($\le500$ Myr) star formation
  history of the centre of SagDIG and an assessment of the MS, BSG and
  RSG  populations  will be  thoroughly analyzed  in  a dedicated
  paper (Held et al. to be submitted).

  The  good  agreement  between  the  {\emph{observed  }  and  {\emph{
        simulated}}     diagrams     for     stars    fainter     than
    $m_{F606W}\simeq23.0$ allows us  to infer that the \ion{H}{i}-hole
    and the comparison  fields are {\emph{similarly}} characterized by
    the onset of significant  star formation between $\sim400-800$ Myr
    ago, for stars basically  populating the red clump. In conclusion,
    the comparison between  {\emph{observed }} and {\emph{ simulated}}
    diagrams provides  a quantitative confirmation for  the absence of
    (Hydrogen  and Helium-burning)  stellar  populations younger  than
    $\sim400$  Myr which  is greater  than  the estimated  age of  the
    SagDIG \ion{H}{i}-hole  ($\sim100$ Myr, \citealt{Momany2005a}) and
    that    in    other     dwarf    galaxies    (see    Table~3    in
    \citealt{Warren2011i}).

    Young, $\lesssim10$  Myr, OB associations  have been traditionally
    (\citealt{Weaver1977u} and  \citealt{Brinks1986c}) associated with
    \ion{H}{i}-hole  (shells/bubble)   formation.  The  {\emph{total}}
    absence  of   OB  associations  in  {\emph{the}}   centre  of  the
    $\sim0.65$   kpc   diameter   SagDIG   \ion{H}{i}-hole   marks   a
    {\emph{distinctive      }}      difference      with      galaxies
    (e.g. \citealt{Kim1999qr} and \citealt{Hatzidimitriou2005} for the
    Magellanic Clouds and \citealt{Weisz2009b} for \ion{Holmberg}{ii})
    displaying   {\emph{multiple}},   \emph{small}}  \ion{H}{i}-holes;
  where     OB    associations    {\emph{were}}     detected    within
  \ion{H}{i}-holes.
  However,  the ``classical'' feedback  theory (and  the search  for a
  single-age   cluster)  {\emph{hardly   }}   provided  a   clear-cut,
  {\emph{undisputed}},   one-to-one  correspondence  between   the  OB
  associations  and the formation  of \ion{H}{i}-holes.  Indeed, cases
  where \ion{H}{i}-holes  ($\sim60$ pc diameter  and $\lesssim10$ Myr)
  had  {\emph{no}} associated stellar  component {\emph{(at  all)} are
    not rare; reaching a  frequency of {\emph{$\sim10\%$}} in the case
    of  the  Small  Magellanic  Cloud  (\citealt{Hatzidimitriou2005}).
    Moreover, the  same study also  established a factor  of $\sim1.5$
    over-abundance of \ion{H}{i}-holes that are {\emph{not}} spatially
    correlated  with an  OB association  compared  to \ion{H}{i}-holes
    that  are,  and surprisingly  infer  similar  properties for  both
    \ion{H}{i}-hole groups.

\begin{figure}
\begin{centering}
\includegraphics[scale=0.165]{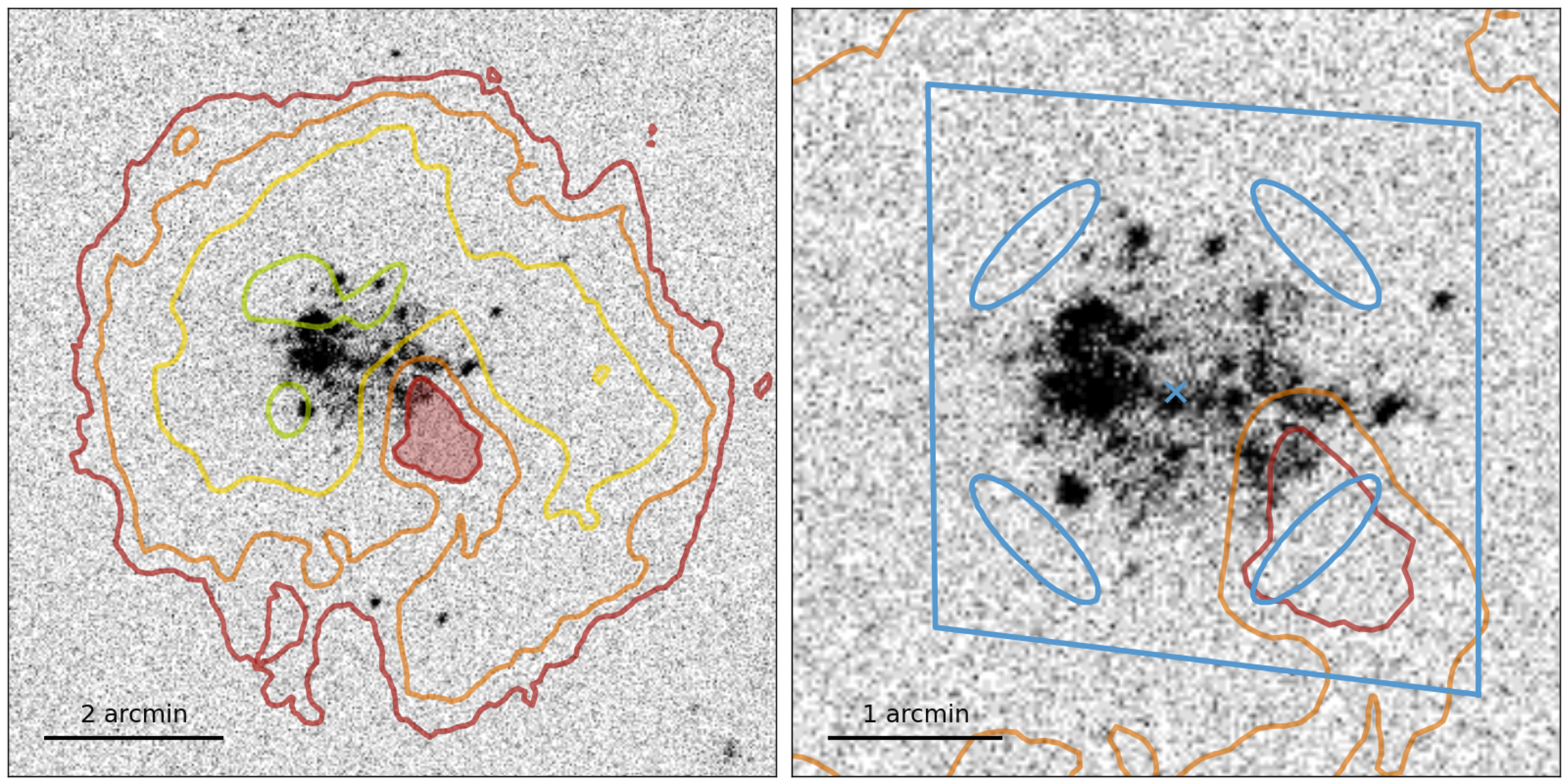}
\par\end{centering}
\centering{}\caption{The two panels display the GALEX far-ultraviolet image of SagDIG upon
which the \ion{H}{i} contours are over-plotted. The 3 highest density
\ion{H}{i} clumps are the inner, open (green) contours, while the
\ion{H}{i}-hole is highlighted as shaded region. The right panel
displays a closer view on the main body of SagDIG highlighting: (i)
the quadrangle ACS@HST field of view, (ii) the SagDIG derived centre
(see text), marked by a cross, and (iii) the $4$ equidistant elliptical
regions; used to extract and compare the stellar populations in the
\ion{H}{i}-hole and the surrounding comparison fields. \label{fig:HI-HOLE}}
\end{figure}

Despite all the reported  difficulties, the classical stellar feedback
hypothesis remains appealing for  the creation of \ion{H}{i}-holes and
shaping  of the  interstellar medium.  Recently, a  ``fine-tuning'' of
this  hypothesis  has  been investigated  (e.g.  \citealt{Weisz2009b},
\citealt{McQuinn2010a},  \citealt{Warren2011i},  \citealt{Cannon2011h}
and  \citealt{Cannon2012q}) because  the search  for  a ``single-age''
stellar  cluster  within   \ion{H}{i}-holes  (by  means  of  estimated
integrated light) has been proved rather an {\emph{ambiguous}} concept
given the  diversity of stellar  populations, of different  ages, that
are  often  detected  within  such  \ion{H}{i}-holes.  In  particular,
allowing  stellar  feedback   to  include  {\emph{mixed-age}}  stellar
populations, there  exists the  possibility that the  re-occurrence of
multiple  supernovae  (spread over  hundreds  of  Myr)  may delay  the
cooling  of  the  \ion{H}{i}   gas,  extending  the  lifetime  of  the
\ion{H}{i}-hole to $\sim0.5$ Gyr (\citealt{Recchi2006a}). Evidence for
this is  found in the  study of \citet{McQuinn2010a},  who reconstruct
the star formation histories of  $20$ dwarf galaxies and estimate that
the  duration of  the starburst  is indeed  around $\sim0.55$  Gyr for
$2/3$  of the their  sample. Of  particular interest  is the  study of
\citet{Warren2011i}  who present  a  detailed analysis  of $5$  nearby
dwarf  irregulars  whose  \ion{H}{i}  morphology  is  dominated  by  a
{\emph{centrally-dominant hole}.
  Following the re-construction of the star formation histories in the
  \ion{H}{i}-holes  for each of  their galaxies,  they then  model and
  quantify the stellar  feedback energy (chemical, spectrophotometric,
  and  stellar energy  evolution) and  compare it  with  the estimated
  energy needed to create  the \ion{H}{i}-holes. Their results clearly
  show that  the integrated stellar  energy well exceeds that  for the
  creation of the holes.
Could stellar feedback from multiple age populations, 
extending to $\sim0.5$ Gyr, save the supernovae hypothesis in SagDIG?

At the beginning  of this section we emphasized  how the proper-motion
decontamination  process  would  allow   us  to  further  explore  the
possibility that  the \ion{H}{i}-hole  is an ancient  structure, older
than the traditionally assumed  age ($\sim100-200$ Myr). Indeed, given
the lack  of a stellar  population younger than $\sim300-400$  Myr, we
have focused our  attention on the possible detection  of a handful of
bright  ($m_{F814W}\sim21.0$)  AGB/red  supergiant stars  (redder  and
distinguishable from  the old red  giant branch) that would  trace the
presence  of a  $\sim1$ Gyr  (now faded)  cluster  whose main-sequence
turnoff  level  is $\sim5$  magnitudes  fainter.  The  \ion{H}{i}-hole
color-magnitude diagram shows no evidence of such stars. Moreover, the
same trend is  seen in the $3$ selected comparison  fields, and one is
led to conclude that  the similarities between the stellar populations
and star formation histories of the \ion{H}{i}-hole and control fields
all point to  {\emph{no }} direct correlation between  the creation of
the  hole  and  the  underlying  recent  ($\lesssim400$  Myr)  stellar
population.
Cases in which stellar feedback (also extended to include multi-age stellar
populations) would have provided the necessary energy to produce
and maintain the \ion{H}{i}-holes also (and always) infer (whenever
control fields were available) that these are indistinguishable from
those in the \ion{H}{i}-holes. {\emph{Regardless of energetic arguments, it will 
always be difficult to attribute a hole in the atomic gas disk to the energetic 
input from stars if the stellar population within the hole is indistinguishable 
from that outside it.}}   

We therefore abandon the  hypothesis that the large \ion{H}{i}-hole in
SagDIG is  related to  the stellar population.  How then, can  a large
hole  form in  the  \ion{H}{i} disk,  offset  from the  centre of  the
galaxy,  in a  region that  has been  largely quiescent  for  the last
$\sim400$ Myr ?
The other possibilities considered in \citet{Momany2005a} for its formation
(top heavy IMF, much more efficient supernovae, presence of molecular
gas, accreted ISM) were all found to be unlikely; but there is a further
possibility not considered in that article. \citet{Wada2000c} showed
not only that kpc scale cavities can form in the ISM of dwarf galaxies
such as the LMC (with almost solid body rotation) as a natural result
of thermal and gravitational instabilities, but that the dispersive
effect of supernovae explosions can actually inhibit the formation
of large structures. Moreover, they found that although the small
scale structure of the ISM was strongly influenced by star formation
(supernovae), that a large-scale filamentary structure evolved even
in models with no star formation at all. Such a process would allow
the formation of a large \ion{H}{i}-hole in SagDIG without the need
to invoke stellar feedback via supernovae.

\begin{figure}
\centering{}\includegraphics[scale=0.45]{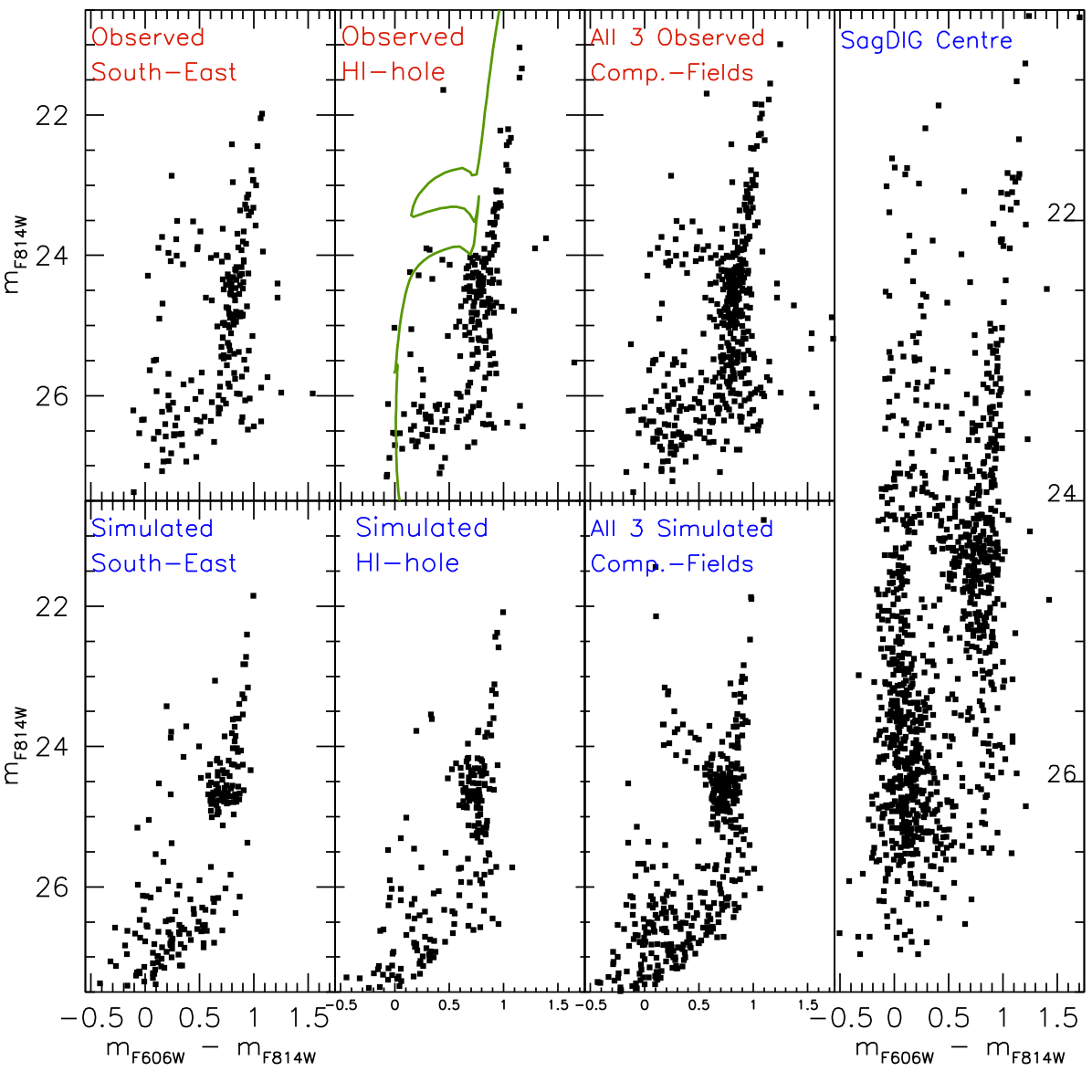}
\caption{From left to right are the color-magnitude
diagrams of the selected (see text) South-East, \ion{H}{i}-hole and
\emph{all} $3$ comparison fields put together. The right-most
panel displays an equal-area elliptical selection around the SagDIG
centre, showing the abundant presence of MS and BSG stellar populations.
Also plotted is a $Z=0.0004$, $0.355$ Gyr old isochrone. 
\label{fig:HI-HOLE-Simulations}}
\end{figure}

The star  formation rate  in SagDIG of  $2.6\times10^{-4}$ M$_{\odot}$
(\citealt{Momany2005a})  implies a  supernovae  rate of  approximately
$2.8\times10^{-6}$ M$_{\odot}$.  This is  $400$ times lower  than that
assumed  for the  LMC by  \citet{Wada2000c}. Given  that  the $V$-band
luminosity of SagDIG is also  approximately $400$ times lower than the
LMC ($M_{V}=-11.5$  compared to $-18.1$), the rate  of energetic input
per unit stellar mass, from supernovae into the ISM of the Sagittarius
dwarf is  similar to that in the  LMC. The atomic gas  mass of SagDIG,
however,  is  only $50$  times  less  than that  of  the  LMC, so  the
energetic  input per  unit atomic  gas mass  is actually  an  order of
magnitude (factor of  $\sim8$) less than in the LMC.  If the effect of
supernovae  is  actually  to  disrupt  the formation  of  large  scale
features  in the  gas, then  SagDIG should  represent a  somewhat more
favorable environment for the formation of large scale features in the
ISM, such as filaments and voids.

Given the lack of evidence or any age stellar population which is
preferentially distributed within the \ion{H}{i}-hole, and the fact
that SagDIG is very isolated (so that there is no obvious source from
which gas may have been acquired externally) we conclude that the
passive formation of a large cavity via gravitational and thermal
instabilities seems to be the most likely scenario for the formation
of the offset \ion{H}{i}-hole in SagDIG.

If gas  disks with little differential rotation,  associated with very
quiescent  stellar  systems,  are   the  ideal  environments  for  the
formation  and  growth  of  filamentary structure  via  the  intrinsic
instability  of the  ISM, then  it may  help explain  otherwise poorly
understood atomic gas morphology in such systems.  One such system may
be the Phoenix dwarf galaxy. In this system the atomic gas is entirely
concentrated in a  curved cloud offset by $590$ pc to  the west of the
stellar  population.  In fact,  if  we were  to  remove  the lowest  2
contours from the \ion{H}{i} map  of SagDIG it would look very similar
to  that  of  Phoenix.   Young  et  al.  (2007)  suggest  this  offset
\ion{H}{i} is caused by  supernovae winds (ram-pressure being excluded
for  lack  of  an  extra-galactic  medium) from  an  episode  of  star
formation  that took  place $\sim100$  Myr ago,  but supernovae-driven
winds  tend to  result in  a symmetrical  gas distribution,  above and
below the disk, in contrast to what is observed. A gas disk that tends
to spontaneously form filamentary  structure on large scales where the
environment  is sufficiently  tranquil (appearing  as `holes'  in maps
with limited spatial resolution) may explain the enigma of these large
holes.

The frequent  lack of a  stellar population that could  be responsible
for  a  given  \ion{H}{i}-hole  has  prompted  \citet{Warren2011i}  to
suggest that  stars alone  may not  be the only  driving force  in the
creation  and maintenance  of \ion{H}{i}-holes,  and  that, ``suitable
local ISM  conditions'', must also prevail.  In  particular, they note
that  star formation  in the  rims of  the large  \ion{H}{i}-holes may
regulate their formation. In this  regard we emphasize that the SagDIG
\ion{H}{i}-hole, unlike most  of those studied by \citet{Warren2011i},
is offset by  $\sim360$ pc from the optical centre  of the galaxy, and
any  recent star  formation episodes  (c.f.  Fig.\,\ref{fig:HI-HOLE}).
In the  context of an intrinsic  instability in the gas,  it is likely
that  any  correlation  between  the border  of  \ion{H}{i}-holes  and
\ion{H}{ii} regions (if present) is  simply a result of star formation
being  promoted in  regions of  greater  gas density;  i.e.  that  gas
morphology drives  the star formation  distribution and not  the other
way round.  } }

\section{Summary \& Conclusions}

A stellar proper-motion study (based on $2$-epoch ACS@HST data-sets)
has allowed us to disentangle the SagDIG stellar populations from
the \emph{heavy} Milky Way foreground contamination. The availability
of cleaned SagDIG \emph{red }stellar populations (being the young
supergiants and intermediate-age asymptotic giant branch stars) permits
an improved examination of the \ion{H}{i}-hole, whose origin
in gas-rich dwarf galaxies is usually assumed to be the result of supernovae
explosions (e.g. \citealt{1998e} and \citealt{Stewart2000bd}). Our
analysis shows that the stellar populations within the \ion{H}{i}-hole
are very \emph{similar} to those outside the hole, associated with $2$ dense 
\ion{H}{i}-clumps. This clearly argues against the \ion{H}{i}-hole being 
the result of supernovae explosions.

As shown previously (Momany et al. 2005), the stellar population 
now present within the hole is consistent with the number of past supernovae 
required to provide the kinetic energy to the ISM to form the hole.   
However, this falls very far short of being a sufficient condition for the
supernovae hypothesis. Regardless of any energetic consideration, the lack of any
distinguishing stellar population within the hole, when compared to
regions outside the hole, argues very strongly against the hole being
caused by energetic input from the stellar population (stellar winds
and supernovae). We find the possibility that gas disks with
solid-body rotation and low levels of star formation can form
large-scale filamentary structure as a result of thermal and
gravitational instability (Wada et al., 2000) to be an appealing
alternative to the supernovae-driven wind hypothesis in SagDIG 
(and likely other similar objects).

As concerning the SagDIG carbon and oxygen-rich samples, previously identified via
the narrow-band filter technique, these have been scrutinized by the proper-motion
membership criterion. Of the original $22$ identified carbon stars,
our analysis proves that $3$ of these are consistent with being Milky
Way thin disk dwarf stars. A population of faint carbon dwarf stars
was identified in high-latitude Sloan Digital Sky Survey fields by
\citet{Downes2004b}, who estimated them to occur with a frequency of a \emph{single}
carbon dwarf per $20$ square degrees. Therefore, the identification
of $3$ Milky Way carbon dwarfs in the SagDIG ACS@HST field of view
(of only $\sim3^{\prime}\times3^{\prime}$), for latitudes of $\sim-16.3^{\circ}$,
is suggestive that the number density of Galactic carbon dwarfs is
quite significant. 

The proper-motion membership criterion also permits the identification
of $7$ $O-$rich stars, $4$ of which display curiously\emph{ red}
colors and \emph{faint} luminosities. Although appealing, there is
no evidence for the presence of oxygen-rich dust enshrouded stars,
and indeed these $4$ SagDIG \emph{red/faint} oxygen-rich stars \emph{do
not} show mid-infrared emission. Their presence however is
a reminder of our poor understanding of the thermal pulsating AGB
phase and the carbon to oxygen-rich transition. Overall, the SagDIG
proper-motion based carbon/oxygen ratio is consistently within the
\citet{Battinelli2013a} reported error. Being the most metal-poor
galaxy in the \citet{Battinelli2013a} sample, SagDIG plays an important
role in establishing the zero-point of the {[}$Fe/H${]}--$log(C/M0+)$
anti-correlation. 

\begin{acknowledgements}
We warmly thank the referee for his/her comments that greatly improved
the presentation of this paper.
\end{acknowledgements}

\bibliographystyle{source/apj}

\newpage{}

{\scriptsize }
\begin{table*}[t]
{\scriptsize }%
\begin{tabular}{|c|c|c|c|c|c|c|c|c|r@{\extracolsep{0pt}.}l|r@{\extracolsep{0pt}.}l|c|}
\hline 
{\scriptsize ID} & {\scriptsize R.A.$_{J2000}$} & {\scriptsize Dec.$_{J2000}$} & {\scriptsize $m_{F475W}$} & {\scriptsize $err._{m_{F475W}}$} & {\scriptsize $m_{F606W}$} & {\scriptsize $err._{m_{F606W}}$} & {\scriptsize $m_{F814W}$} & {\scriptsize $err._{m_{F814W}}$} & \multicolumn{2}{c|}{{\scriptsize $\triangle X_{pixel}$}} & \multicolumn{2}{c|}{{\scriptsize $\triangle Y_{pixel}$}} & {\footnotesize Type}\tabularnewline
\hline 
\hline 
{\scriptsize 1779} & {\scriptsize 292.470424223} & {\scriptsize --17.694280556} & {\scriptsize 24.409} & {\scriptsize 0.008} & {\scriptsize 22.616} & {\scriptsize 0.005} & {\scriptsize 20.762} & {\scriptsize 0.009} & {\scriptsize 0}&{\scriptsize 010} & {\scriptsize --0}&{\scriptsize 025} & {\scriptsize carbon {[}B\&D{]}}\tabularnewline
\hline 
{\scriptsize 8477} & {\scriptsize 292.484743893} & {\scriptsize --17.679781146} & {\scriptsize 23.825} & {\scriptsize 0.006} & {\scriptsize 22.212} & {\scriptsize 0.006} & {\scriptsize 20.454} & {\scriptsize 0.012} & {\scriptsize 0}&{\scriptsize 008} & {\scriptsize --0}&{\scriptsize 021} & {\scriptsize carbon {[}B\&D{]}}\tabularnewline
\hline 
{\scriptsize 11906} & {\scriptsize 292.489788088} & {\scriptsize --17.678259855} & {\scriptsize 23.819} & {\scriptsize 0.006} & {\scriptsize 22.055} & {\scriptsize 0.005} & {\scriptsize 20.359} & {\scriptsize 0.012} & {\scriptsize --0}&{\scriptsize 001} & {\scriptsize{} 0}&{\scriptsize 001} & {\scriptsize carbon {[}B\&D{]}}\tabularnewline
\hline 
{\scriptsize 22018} & {\scriptsize 292.500279743} & {\scriptsize --17.693105356} & {\scriptsize 24.088} & {\scriptsize 0.010} & {\scriptsize 21.996} & {\scriptsize 0.007} & {\scriptsize 20.428} & {\scriptsize 0.010} & {\scriptsize 0}&{\scriptsize 013} & {\scriptsize --0}&{\scriptsize 005} & {\scriptsize carbon {[}B\&D{]}}\tabularnewline
\hline 
{\scriptsize 28484} & {\scriptsize 292.508378693} & {\scriptsize --17.701596633} & {\scriptsize 23.606} & {\scriptsize 0.006} & {\scriptsize 21.647} & {\scriptsize 0.004} & {\scriptsize 19.797} & {\scriptsize 0.011} & {\scriptsize 0}&{\scriptsize 025} & {\scriptsize --0}&{\scriptsize 012} & {\scriptsize carbon {[}B\&D{]}}\tabularnewline
\hline 
{\scriptsize 28552} & {\scriptsize 292.506976257} & {\scriptsize --17.685159138} & {\scriptsize 25.278} & {\scriptsize 0.014} & {\scriptsize 22.784} & {\scriptsize 0.008} & {\scriptsize 20.658} & {\scriptsize 0.012} & {\scriptsize 0}&{\scriptsize 005} & {\scriptsize --0}&{\scriptsize 009} & {\scriptsize carbon {[}B\&D{]}}\tabularnewline
\hline 
{\scriptsize 36807} & {\scriptsize 292.521527421} & {\scriptsize --17.695199055} & {\scriptsize 24.143} & {\scriptsize 0.007} & {\scriptsize 22.132} & {\scriptsize 0.005} & {\scriptsize 20.327} & {\scriptsize 0.009} & {\scriptsize 0}&{\scriptsize 004} & {\scriptsize --0}&{\scriptsize 011} & {\scriptsize carbon {[}B\&D{]}}\tabularnewline
\hline 
{\scriptsize 35768} & {\scriptsize 292.517402245} & {\scriptsize --17.675026482} & {\scriptsize 23.763} & {\scriptsize 0.005} & {\scriptsize 21.989} & {\scriptsize 0.004} & {\scriptsize 20.276} & {\scriptsize 0.009} & {\scriptsize --0}&{\scriptsize 006} & {\scriptsize --0}&{\scriptsize 006} & {\scriptsize carbon {[}Cook+B\&D{]}   }\tabularnewline
\hline 
{\scriptsize 26439} & {\scriptsize 292.504205982} & {\scriptsize --17.681264796} & {\scriptsize 23.578} & {\scriptsize 0.008} & {\scriptsize 21.657} & {\scriptsize 0.008} & {\scriptsize 19.883} & {\scriptsize 0.014} & {\scriptsize 0}&{\scriptsize 006} & {\scriptsize --0}&{\scriptsize 004} & {\scriptsize carbon {[}Cook+B\&D{]}  }\tabularnewline
\hline 
{\scriptsize 23543} & {\scriptsize 292.501195979} & {\scriptsize --17.684096543} & {\scriptsize 24.799} & {\scriptsize 0.013} & {\scriptsize 22.175} & {\scriptsize 0.009} & {\scriptsize 20.069} & {\scriptsize 0.018} & {\scriptsize 0}&{\scriptsize 010} & {\scriptsize --0}&{\scriptsize 022} & {\scriptsize carbon {[}Cook+B\&D{]}  }\tabularnewline
\hline 
{\scriptsize 2246} & {\scriptsize 292.470442998} & {\scriptsize --17.675733036} & {\scriptsize 23.773} & {\scriptsize 0.004} & {\scriptsize 21.775} & {\scriptsize 0.005} & {\scriptsize 19.923} & {\scriptsize 0.011} & {\scriptsize 0}&{\scriptsize 001} & {\scriptsize -0}&{\scriptsize 026} & {\scriptsize carbon {[}Cook+B\&D{]}   }\tabularnewline
\hline 
{\scriptsize 36305} & {\scriptsize 292.517750836} & {\scriptsize --17.665476415} & {\scriptsize 23.867} & {\scriptsize 0.004} & {\scriptsize 22.749} & {\scriptsize 0.004} & {\scriptsize 21.573} & {\scriptsize 0.006} & {\scriptsize --0}&{\scriptsize 003} & {\scriptsize{} 0}&{\scriptsize 002} & {\scriptsize carbon {[}Cook{]}}\tabularnewline
\hline 
{\scriptsize 29836} & {\scriptsize 292.50700088} & {\scriptsize --17.667797142} & {\scriptsize 22.797} & {\scriptsize 0.005} & {\scriptsize 21.100} & {\scriptsize 0.006} & {\scriptsize 19.570} & {\scriptsize 0.014} & {\scriptsize 0}&{\scriptsize 005} & {\scriptsize --0}&{\scriptsize 002} & {\scriptsize carbon {[}Cook{]}}\tabularnewline
\hline 
{\scriptsize 26867} & {\scriptsize 292.504035325} & {\scriptsize --17.674187250} & {\scriptsize 23.291} & {\scriptsize 0.005} & {\scriptsize 21.878} & {\scriptsize 0.005} & {\scriptsize 20.564} & {\scriptsize 0.012} & {\scriptsize 0}&{\scriptsize 008} & {\scriptsize --0}&{\scriptsize 011} & {\scriptsize carbon {[}Cook{]}}\tabularnewline
\hline 
{\scriptsize 23532} & {\scriptsize 292.500977236} & {\scriptsize --17.681787246} & {\scriptsize 23.989} & {\scriptsize 0.010} & {\scriptsize 22.259} & {\scriptsize 0.008} & {\scriptsize 20.555} & {\scriptsize 0.011} & {\scriptsize 0}&{\scriptsize 013} & {\scriptsize -0}&{\scriptsize 028} & {\scriptsize carbon {[}Cook{]}}\tabularnewline
\hline 
{\scriptsize 22225} & {\scriptsize 292.499001740} & {\scriptsize --17.676113300 } & {\scriptsize 23.264} & {\scriptsize 0.007} & {\scriptsize 21.493} & {\scriptsize 0.006} & {\scriptsize 19.830} & {\scriptsize 0.013} & {\scriptsize 0}&{\scriptsize 002} & {\scriptsize --0}&{\scriptsize 013} & {\scriptsize carbon {[}Cook{]}}\tabularnewline
\hline 
{\scriptsize 15566} & {\scriptsize 292.494754608} & {\scriptsize --17.685898547} & {\scriptsize 24.029} & {\scriptsize 0.010} & {\scriptsize 22.948} & {\scriptsize 0.007} & {\scriptsize 21.903} & {\scriptsize 0.007} & {\scriptsize 0}&{\scriptsize 006} & {\scriptsize{} 0}&{\scriptsize 007} & {\scriptsize carbon {[}Cook{]}}\tabularnewline
\hline 
{\scriptsize 6945} & {\scriptsize 292.480575696} & {\scriptsize --17.666244664} & {\scriptsize 23.725} & {\scriptsize 0.006} & {\scriptsize 22.667} & {\scriptsize 0.004} & {\scriptsize 21.537} & {\scriptsize 0.006} & {\scriptsize 0}&{\scriptsize 002} & {\scriptsize --0}&{\scriptsize 011} & {\scriptsize carbon {[}Cook{]}}\tabularnewline
\hline 
{\scriptsize 6090} & {\scriptsize 292.479789823} & {\scriptsize --17.677937700  } & {\scriptsize 23.355} & {\scriptsize 0.005} & {\scriptsize 21.475} & {\scriptsize 0.007} & {\scriptsize 19.786} & {\scriptsize 0.015} & {\scriptsize 0}&{\scriptsize 003} & {\scriptsize -0}&{\scriptsize 030} & {\scriptsize carbon {[}Cook{]}}\tabularnewline
\hline 
{\scriptsize 2706} & {\scriptsize 292.470758949} & {\scriptsize --17.662969603} & {\scriptsize 24.733} & {\scriptsize 0.010} & {\scriptsize 23.325} & {\scriptsize 0.007} & {\scriptsize 21.725} & {\scriptsize 0.007} & {\scriptsize --0}&{\scriptsize 181} & {\scriptsize --0}&{\scriptsize 192} & {\scriptsize Galactic carbon dwarf   }\tabularnewline
\hline 
{\scriptsize 22924} & {\scriptsize 292.499538712} & {\scriptsize --17.673751215} & {\scriptsize 24.461} & {\scriptsize 0.008} & {\scriptsize 23.082} & {\scriptsize 0.006} & {\scriptsize 21.734} & {\scriptsize 0.007} & {\scriptsize --0}&{\scriptsize 086} & {\scriptsize --0}&{\scriptsize 033} & {\scriptsize Galactic carbon dwarf   }\tabularnewline
\hline 
{\scriptsize 6491} & {\scriptsize 292.480173480} & {\scriptsize --17.672926541} & {\scriptsize 23.835} & {\scriptsize 0.006} & {\scriptsize 21.884} & {\scriptsize 0.005} & {\scriptsize 20.095} & {\scriptsize 0.012} & {\scriptsize 0}&{\scriptsize 021} & {\scriptsize --0}&{\scriptsize 294} & {\scriptsize Galactic carbon dwarf  }\tabularnewline
\hline 
{\scriptsize 35304} & {\scriptsize 292.517603956} & {\scriptsize --17.688463174} & {\scriptsize 23.502 } & {\scriptsize 0.006} & {\scriptsize 22.209} & {\scriptsize 0.005} & {\scriptsize 20.890} & {\scriptsize 0.006} & {\scriptsize 0}&{\scriptsize 003} & {\scriptsize --0}&{\scriptsize 006} & {\scriptsize oxygen {[}B\&D{]}}\tabularnewline
\hline 
{\scriptsize 36315} & {\scriptsize 292.517452061} & {\scriptsize --17.661983977} & {\scriptsize 22.921} & {\scriptsize 0.004} & {\scriptsize 21.505} & {\scriptsize 0.003} & {\scriptsize 19.604} & {\scriptsize 0.014} & {\scriptsize 0}&{\scriptsize 035 } & {\scriptsize --0}&{\scriptsize 013} & {\scriptsize oxygen {[}B\&D{]}}\tabularnewline
\hline 
{\scriptsize 32223} & {\scriptsize 292.512707052} & {\scriptsize --17.698337255} & {\scriptsize 25.005} & {\scriptsize 0.009} & {\scriptsize 23.437} & {\scriptsize 0.006} & {\scriptsize 21.322} & {\scriptsize 0.006} & {\scriptsize 0}&{\scriptsize 016} & {\scriptsize 0}&{\scriptsize 011} & {\scriptsize oxygen {[}B\&D{]}}\tabularnewline
\hline 
{\scriptsize 24461} & {\scriptsize 292.504056970} & {\scriptsize --17.704380499} & {\scriptsize 26.152} & {\scriptsize 0.019} & {\scriptsize 24.327} & {\scriptsize 0.007} & {\scriptsize 21.478} & {\scriptsize 0.007} & {\scriptsize 0}&{\scriptsize 021} & {\scriptsize --0}&{\scriptsize 029} & {\scriptsize oxygen {[}B\&D{]}}\tabularnewline
\hline 
{\scriptsize 8170} & {\scriptsize 292.482237066} & {\scriptsize --17.658999926} & {\scriptsize 23.050} & {\scriptsize 0.006} & {\scriptsize 21.691} & {\scriptsize 0.007} & {\scriptsize 20.224} & {\scriptsize 0.010} & {\scriptsize --0}&{\scriptsize 011} & {\scriptsize 0}&{\scriptsize 008} & {\scriptsize oxygen {[}B\&D{]}}\tabularnewline
\hline 
{\scriptsize 3971} & {\scriptsize 292.473941125} & {\scriptsize --17.663168733} & {\scriptsize 25.209 } & {\scriptsize 0.014} & {\scriptsize 23.762} & {\scriptsize 0.008} & {\scriptsize 21.612} & {\scriptsize 0.010} & {\scriptsize --0}&{\scriptsize 018} & {\scriptsize --0}&{\scriptsize 027} & {\scriptsize oxygen {[}B\&D{]}}\tabularnewline
\hline 
{\scriptsize 2577} & {\scriptsize 292.470892782} & {\scriptsize --17.668448303} & {\scriptsize 25.747} & {\scriptsize 0.011} & {\scriptsize 24.278} & {\scriptsize 0.009} & {\scriptsize 21.856} & {\scriptsize 0.006} & {\scriptsize 0}&{\scriptsize 007} & {\scriptsize --0}&{\scriptsize 025} & {\scriptsize oxygen {[}B\&D{]}}\tabularnewline
\hline 
\end{tabular}{\scriptsize \par}

{\scriptsize \caption{The SagDIG photometric ($m_{F475W}$, $m_{F606W}$, and $m_{F814W}$
and their relative errors) and astrometric (J2000 coordinates and
the pixel-based offset between the two $F814W$ epochs) properties
of the known carbon and oxygen stars in our ACS@HST central pointing.
The last column reports the identification reference. The entire SagDIG
catalog is made available in its entirety (in the same format as in
this table) via the link to the machine-readable version above.\label{tab:C-stars-1}}
}{\scriptsize \par}

\end{table*}
{\scriptsize \par}

\newpage{}

\end{document}